\documentclass[12pt]{article}
\usepackage{amssymb}
\usepackage{epsfig}
\usepackage{bbm}
\usepackage{bbold}
\usepackage{graphicx}
\usepackage{hyperref}
\usepackage{romannum}

\newcommand\pubnumber{Pasquale Di Bari, NuPhys2018}
\newcommand\pubdate{\today}

\def\soton{School of Physics and Astronomy, University of Southampton, \\
Southampton, SO17 1BJ, UK}
\def\support{\footnote{PDB  acknowledges financial support from the STFC Consolidated Grant L000296/1. 
This project has received funding/support from the European Union Horizon 2020 research and innovation 
programme under the Marie Sk\l{}'odowska-Curie grant agreements number 690575 and  674896.}}

\def\Title#1{\begin{center} {\Large #1 } \end{center}}
\def\Author#1{\begin{center}{ \sc #1} \end{center}}
\def\Address#1{\begin{center}{ \it #1} \end{center}}

\newcommand\pubblock{\rightline{\begin{tabular}{l} \pubnumber\\
         \pubdate  \end{tabular}}}
\newenvironment{Abstract}{\begin{quotation}  }{\end{quotation}}
\newenvironment{Presented}{\begin{quotation} \begin{center} 
             PRESENTED AT\end{center}\bigskip 
      \begin{center}\begin{large}}{\end{large}\end{center} \end{quotation}}
\def\Acknowledgements{\bigskip  \bigskip \begin{center} \begin{large}
             \bf ACKNOWLEDGEMENTS \end{large}\end{center}}

\def\beq{\begin{equation}}
\def\eeq#1{\label{#1}\end{equation}}
\def\eeqn{\end{equation}}

\def\beqa{\begin{eqnarray}}
\def\eeqa#1{\label{#1}\end{eqnarray}}
\def\eeqan{\end{eqnarray}}

\let\bar=\overbar

\def\VEV#1{\left\langle{ #1} \right\rangle}

\def\D{{\cal D}}
\def\L{{\cal L}}

\def\O{{\cal O}}

\def\Dslash{\not{\hbox{\kern-4pt $D$}}}
\def\dslash{\not{\hbox{\kern-2pt $\del$}}}

\def\ee{e^+e^-}

\def\mt{m_t}

\def\msb{{\bar{\ssstyle M \kern -1pt S}}}

\def\s#1{\widetilde{#1}}

\begin{document}

\def\a{\alpha}
\def\b{\beta}
\def\c{\chi}
\def\d{\delta}
\def\e{\epsilon}
\def\f{\phi}
\def\g{\gamma}
\def\h{\eta}
\def\i{\iota}
\def\j{\psi}
\def\k{\kappa}
\def\la{\lambda}
\def\m{\mu}
\def\n{\nu}
\def\o{\omega}
\def\p{\pi}
\def\q{\theta}
\def\r{\rho}
\def\s{\sigma}
\def\t{\tau}
\def\u{\upsilon}
\def\x{\xi}
\def\z{\zeta}
\def\D{\Delta}
\def\F{\Phi}
\def\G{\Gamma}
\def\J{\Psi}
\def\L{\Lambda}
\def\O{\Omega}
\def\P{\Pi}
\def\Q{\Theta}
\def\S{\Sigma}
\def\U{\Upsilon}
\def\X{\Xi}

\def\ve{\varepsilon}
\def\vf{\varphi}
\def\vr{\varrho}
\def\vs{\varsigma}
\def\vq{\vartheta}

\def\dg{\dagger}                                     
\def\ddg{\ddagger}                                   
\def\wt#1{\widetilde{#1}}                    
\def\mt{\widetilde{m}_1}
\def\mti{\widetilde{m}_i}
\def\rt{\widetilde{r}_1}
\def\mtt{\widetilde{m}_2}
\def\mttt{\widetilde{m}_3}
\def\rtt{\widetilde{r}_2}
\def\mb{\overline{m}}
\def\VEV#1{\left\langle #1\right\rangle}        
\def\be{\begin{equation}}
\def\ee{\end{equation}}
\def\ds{\displaystyle}
\def\ra{\rightarrow}

\def\bea{\begin{eqnarray}}
\def\eea{\end{eqnarray}}
\def\NO{\nonumber}
\def\Bar#1{\overline{#1}}

\begin{titlepage}
\pubblock

\vfill
\Title{Neutrino masses, leptogenesis and dark matter}
\vfill
\Author{ Pasquale Di Bari\support}
\Address{\soton}
\vfill
\pagenumbering{arabic}
\begin{Abstract}
Despite the lack of evidence of new physics at colliders, neutrino masses, dark matter and
matter-antimatter asymmetry of the universe require an extension of the standard model. 
After discussing some new concepts and tools in the description
of seesaw neutrino models, such as motion in flavour lepton space and the introduction of 
the bridging matrix, that nicely complements the orthogonal matrix,
I discuss recent developments in the connections between neutrino data and scenarios of leptogenesis 
within some well motivated extensions of the standard model. 
Finally, I briefly review a simple unified picture of neutrino masses, leptogenesis and dark matter
based on an extension of the seesaw Lagrangian 
where a non-renormalizable effective operator coupling right-handed neutrinos to the Higgs
is introduced. This operator enhances the right-handed neutrino mixing 
between a coupled heavy right-handed neutrino and a decoupled one playing the role of dark matter.
Moreover the interference between the two coupled RH neutrinos also generates 
$C\!P$ violation and this allows to achieve successful leptogenesis.
Although the dark matter right-handed neutrino escapes direct and collider searches, 
its decays produce a detectable contribution to the very high energy neutrino 
flux discovered by IceCube, making the model predictive and testable at neutrino telescopes. 
\end{Abstract}
\vfill
\begin{Presented}
NuPhys 2018: Prospects in Neutrino Physics \\
Cavendish Centre, London, 19-21 December, 2018
\end{Presented}
\vfill
\end{titlepage}
\def\thefootnote{\fnsymbol{footnote}}
\setcounter{footnote}{0}

\section{Introduction}
Although there is no evidence of new physics at colliders,\footnote{Barring of course flavour anomalies, 
currently lacking both a consistent and sufficiently statistically significant 
experimental support to be regarded as evidences.}
there are clear phenomenological
reasons to extend the standard model (SM): an explanation of neutrino masses and mixing, understanding
the nature of dark matter (DM) and the origin of the matter-antimatter asymmetry of the universe. 
It is reasonable that these three compelling evidences of new physics should be addressed within a unified 
picture. We discuss how a simple extension of the SM based on the introduction 
of right-handed neutrinos with Yukawa couplings, a Majorana mass term and additional couplings 
to the Higgs described by a non-renormalisable effective operator, allows to address all three of them. 
Neutrino masses and mixing are explained by a simple
type-I seesaw mechanism, matter anti-matter asymmetry of the universe by leptogenesis, 
while a decoupled right-handed neutrino can play the role of DM. The picture also predicts a contribution to 
the very high energy neutrino flux that makes it testable at neutrino telescopes. The recent
discovery by the IceCube detector is then extremely interesting and provides a way to test this unified model.

Seesaw neutrino models and scenarios of leptogenesis are  nicely embeddable within models of new physics,
in particular grand-unified models, models of flavour and combinations of both. I will discuss in general
how these models can point to different regions in the  seesaw parameter space that, especially thanks to leptogenesis, can imply different predictions on low energy neutrino parameters. From this point of view, I will mainly emphasise the  importance of absolute neutrino mass scale experiments. 
I will also discuss some recent developments in representing seesaw neutrino models in lepton flavour space and how the introduction of a new matrix, the bridging matrix, nicely complements the orthogonal matrix in describing
and understanding seesaw neutrino models.

\section{Neutrino masses and mixing}

Neutrino mixing experiments measure two neutrino mass squared differences that can be
expressed in terms of the atmospheric neutrino mass scale, 
$m_{\rm atm} \equiv \sqrt{|m^2_3 - m^2_1|} = (49.9 \pm 0.3)\,{\rm meV}$, 
and the solar neutrino mass scale, 
$m_{\rm sol} \equiv \sqrt{m^2_2 - m^2_1} = (8.6 \pm 0.1)\,{\rm meV}$,
where the numerical values are the results of a recent global analysis \cite{nufit2018}. 
This necessarily implies that for $m_1 \gg m_{\rm atm}$ all three neutrino masses are quasi-degenerate,
while for $m_1 \ll m_{\rm sol}$ one obtains either normal or  inverted 
hierarchical limit, depending on the sign of $m^2_3 - m^2_1$ that is still unknown. 
If we indicate with $m_{1'}$ the lightest neutrino mass, coinciding with $m_1$ in the case of normal
ordering, for positive  $m^2_3 - m^2_1$, and with $m_3$ in the case of inverted ordering, for 
negative $m^2_3 - m^2_1$, cosmological observations combined with 
neutrino mixing results, place the most stringent upper bound $m_{1'} \lesssim 0.07\,{\rm eV} \, (95 \% \, {\rm CL})$ excluding 
quasi-degenerate neutrinos
\cite{planck2015}.\footnote{The latest results from the {\em Planck} collaboration \cite{planck2018}, place an upper bound on the sum of neutrino masses $\sum_i \, m_i \lesssim 0.12\,{\rm eV} \, (95\% {\rm CL})$. 
Combining this upper bound with the neutrino mixing results, this implies 
$m_{1'}\lesssim 0.03\,(0.017)\,{\rm eV}$ for normal (inverted) ordering. It should  also be noticed that
inverted ordering is disfavoured at $\sim 1.6\,\sigma$. This goes in the same direction
of latest results from neutrino mixing experiments also disfavouring normal ordering as mentioned in the text.}

In the case of normal ordering, favoured over inverted ordering at $\sim 3\s$,
latest global analyses find the following $3\sigma$ ranges 
for the mixing angles and Dirac phase in the leptonic mixing matrix $U$ 
(switching from mass to weak eigenstates such that $\nu_\a = U_{\a i}\,\nu_i$) \cite{nufit2018} 
\begin{eqnarray}
\theta_{12} & = &  32^\circ \, \mbox{-- } 36^\circ \,   ,  \\ \nonumber
\theta_{13} & = &  8.2^\circ \, \mbox{-- } 9.0^\circ \,   ,  \\  \nonumber
\theta_{23} & = &  41^\circ \, \mbox{-- } 52^\circ \,   ,  \\     \nonumber
\delta & = & 135^\circ \, \mbox{-- } 366^\circ \,  .
\end{eqnarray}
An easy (and economical) way to extend the SM to account for neutrino masses and mixing
is to introduce  right-handed (RH) neutrinos with Yukawa couplings $h^{\nu}$ to left-handed (LH) neutrinos and Higgs
so that, in the flavour basis where the Yukawa charged lepton matrix is diagonal,  
the charged lepton and neutrino Yukawa interactions can be written as
\begin{equation} \label{LYM}
- {\cal L}_{Y}^{\nu + \ell} =\overline{L_{\a}}\,h^{\ell}_{\a\a}\,{\a}_{R }\, \Phi 
+ \overline{L_{\a}}\,h^{\nu}_{\a J}\,N_{R J}\, \widetilde{\Phi}   \,  ,
\end{equation}
where $L^T_\a \equiv (\nu_{L\a},\a_L)$ are the leptonic doublets, $M_\Romannum{1} \leq \dots \leq M_N$ are the heavy neutrino masses  and we indicate with Greek indexes, $\a=e,\mu,\tau$,
the charged lepton flavour, and with Roman indexes, $J ={\Romannum{1}}, {\Romannum{2}} \dots, N$, 
the heavy neutrino flavour. For definiteness we can consider the simple and well motivated case $N =3$. 
After electroweak spontaneous symmetry breaking the Higgs vev generates a neutrino Dirac mass matrix
$m_{D} =v\,h^\nu$ and charged lepton masses $m_{\a} =v\,h^{\ell}_{\a\a}$.  
In this way the leptonic mass term in the Lagrangian for neutrinos and charged leptons reads
\be\label{massterm}
- {\cal L}^{\ell + \nu}_{m} = \overline{{\a}_{L}}\, m_\a \,{\a}_{R} +  
\overline{\nu_{L \a }}\, m_{D \a J} \,  N_{R J} \,  .
\ee
If we now operate a bi-unitary transformation 
of the LH and RH neutrino fields, $N_{Ri} = U_{R\, iJ}\,N_{R J}$ and $\nu_{L i}=V_{L i \a}\, \nu_{L\a}$ ($i=1,2,3$),
that switches to the Yukawa basis where the neutrino Dirac mass matrix is diagonal,
one obtains a bi-unitary parameterisation of the neutrino Dirac mass matrix 
(mathematically, its singular value decomposition) 
\be\label{biunitary}
m_D = V^\dagger_L \, D_{m_D} \, U_R \,  ,
\ee  
where $D_{m_D} \equiv {\rm diag}(m_{D1}, m_{D2}, m_{D3})$   and $m_{D1} < m_{D2} < m_{D3}$
are the Dirac neutrino masses. In this case  one has that the Yukawa basis coincides with the
neutrino mass eigenstate basis. The neutrino masses simply coincide with the neutrino Dirac masses, 
$m_i = m_{D i}$ and the leptonic mixing matrix is given by $U = V^\dagger_L$.

However, this simple option, by itself, does not address the cosmological puzzles and 
also implies that neutrino Yukawa couplings are much smaller than those of the other massive
fermions, especially when the comparison is made within the same family, so that,
for example, $m_3/m_{\rm top} \sim 10^{-12}$.\footnote{This does not mean that Dirac neutrinos 
cannot emerge within well motivated models, but these involve additional ingredients
compared to the minimal extension we are considering so that
the picture is actually less minimal that one can think. For example, ways to justify 
the lightness of Dirac neutrinos are found within frameworks with large \cite{arkani} 
or warped \cite{neubert} extra-dimensions. Also a mechanism of leptogenesis with Dirac
neutrinos has been proposed \cite{lindner} but it still requires some 
external source, such as GUT baryogenesis, of an initial $B+L$ asymmetry.}

However, if in addition to RH neutrinos and a Yukawa coupling term, one also allows for lepton number violation, considering that neutrinos are neutral, then in general one also has a right-right Majorana mass term, so that, after electroweak spontaneous symmetry breaking,  one obtains for the leptonic mass term (\ref{massterm}) 
\be\label{seesawmasster}
- {\cal L}^{\ell + \nu}_{m} = \overline{{\a}_{L}}\, m_\a \,{\a}_{R} +  
\overline{\nu_{L \a }}\, m_{D \a J} \,  N_{R J} +
{1\over 2}\,\overline{N_{R J}^{\,c}}\, M_J\, N_{R J} + {\rm h.c.}  \,  ,
\ee
in a basis where both charged lepton Dirac mass matrix and Majorana mass matrix are diagonal
(we will refer to it as the {\em flavour} basis) and where $M_{\Romannum{1}} \leq M_{\Romannum{2}} \leq M_{\Romannum{3}}$.   In the seesaw limit, for $M_\Romannum{1} \gg m_{D 3}$, the spectrum of mass
eigenstates splits into a heavy set, with masses basically coinciding with the three $M_{J}$'s
and the eigenstates with the RH neutrinos, and into a light set with the mass weigenstates
almost coinciding with the LH neutrinos and with masses given by the seesaw formula \cite{seesaw}
\be\label{seesaw}
m_i = U^{\ast}_{i \a} \, m_{D \a J} \, M^{-1}_{J} \, (m_D^T)_{J \b} \, U^{\ast}_{\b i} \,  .
\ee
Neutrinos are predicted to be Majorana neutrinos and, therefore, in the parameterisation of the leptonic mixing matrix, one has also to include two Majorana phases that cannot be reabsorbed in
the fields and can in principle be measured in new physical processes involving 
lepton number violation. In practice, currently, we can just hope to observe neutrinoless
double beta decay that would correspond to measure just one Majorana phase. 

The seesaw formula is equivalent to the orthogonality of the matrix \cite{casasibarra}
\be\label{orthogonal}
\O_{i J} = {(U^\dagger \, m_{D})_{i J} \over \sqrt{m_i \, M_J}} \,  ,
\ee
providing a useful ({\em orthogonal}) parameterisation of the neutrino Dirac mass matrix 
\be\label{orthogonal}
m_{D \a J} = U_{\a i}\,\sqrt{m_i} \, \O_{i J}\,\sqrt{M_J} \,  .
\ee
The orthogonal matrix elements provide the  fractional contribution to the light neutrino mass $m_i$ from the term proportional to the inverse heavy neutrino mass $M_J^{-1}$ and, very importantly, they tell how fine-tuned are phase  cancellations in the seesaw formula to get each  $m_i$ as a sum of  terms $ \propto M_J^{-1}$.

Another interesting matrix,  
useful in the study of the seesaw neutrino mass models  is the {\em bridging matrix}, 
recently introduced in \cite{pdbrefsam}, defined as
\be\label{XiJ}
B_{i J} \equiv {(U^\dagger \, m_D)_{i J} \over \sqrt{(m_D^\dagger \, m_D)_{JJ}}}   \,  .
\ee
This operates the transformation between the lepton flavour basis  determined by the neutrino mass eigenstates  to that one determined by heavy neutrino lepton flavour states (and vice-versa). If one considers lepton doublet  
states, one has indeed $|L_J\rangle = B_{iJ}\,|L_i\rangle$ (or considering lepton doublet fields one has $L_J = (B^\dagger)_{Ji} \, L_i$).  

The orthogonal parameterisation (\ref{orthogonal}) clearly shows that within a standard seesaw extension of the SM with three RH neutrinos, eighteen new parameters are introduced: nine low energy neutrino parameters 
and nine high energy neutrino parameters that escape completely the information from laboratory experiments.
In particular the mass spectrum of the three heavy (mostly RH) neutrinos is one of the main unknowns. From this point of view, as we will discuss in the next section, leptogenesis plays an important role in connecting the
heavy neutrino mass spectrum to the matter-antimatter asymmetry of the universe. 

However, from the  bi-unitary parameterisation of the neutrino
Dirac mass matrix Eq.~(\ref{biunitary}), one can get important insight 
in the the different classes of seesaw neutrino models that 
can be built. If we plug the bi-unitary parameterisation  into the seesaw formula (\ref{seesaw}),
this can be written as
\be\label{seesawbiunitary}
m_{\nu}\equiv U\,D_m \, U^T = V^\dagger_L \, D_{m_D} \, U_R \, D_M^{-1} \, U_R^T \, D_{m_D} \, V^\ast_L  \,  ,
\ee
where $m_{\nu}$ is the light neutrino
mass matrix (in the weak basis), $D_m \equiv {\rm diag}(m_1, m_2, m_3)$  and $D_M \equiv 
{\rm diag}(M_{\Romannum{1}},M_{\Romannum{2}},M_{\Romannum{3}})$.

\subsubsection*{All mixing from LH sector: form-dominance models}

We have seen that in the absence of a Majorana mass term, all the mixing is described by the 
mismatch in the LH sector between the Yukawa basis and the charged lepton flavour basis
so that $U=V_L^\dagger$. In the presence of a Majorana mass term this is still
a perfectly viable possibility. This means that in this case Majorana mass 
and  neutrino Dirac mass matrices are diagonal in the same basis and $U_R = I$. 
Eq.~(\ref{seesawbiunitary}) then immediately confirms that $U = V_L^\dagger$,
as in the case of Dirac neutrinos, but this time one obtains seesawed neutrino masses 
$m_i = m^2_{Dj}/M_J$. 

In these models  
the orthogonal matrix $\O$ is simply given by the permutation matrix and the fine-tuning is 
minimal \cite{geometry} and they are usually referred to as 
{\em form-dominance models} \cite{chenking}. 
The three flavour bases of light neutrino mass eigenstates, Yukawa basis 
and heavy neutrino flavour basis coincide. In the figure we show in lepton flavour space,
using the light neutrino mass eigenstate as reference basis, the weak basis (left panel) and 
the heavy neutrino flavour basis (right panel) in the case that this coincides with light neutrino basis.
\begin{figure}[htb]
\centering
\includegraphics[height=1.7in]{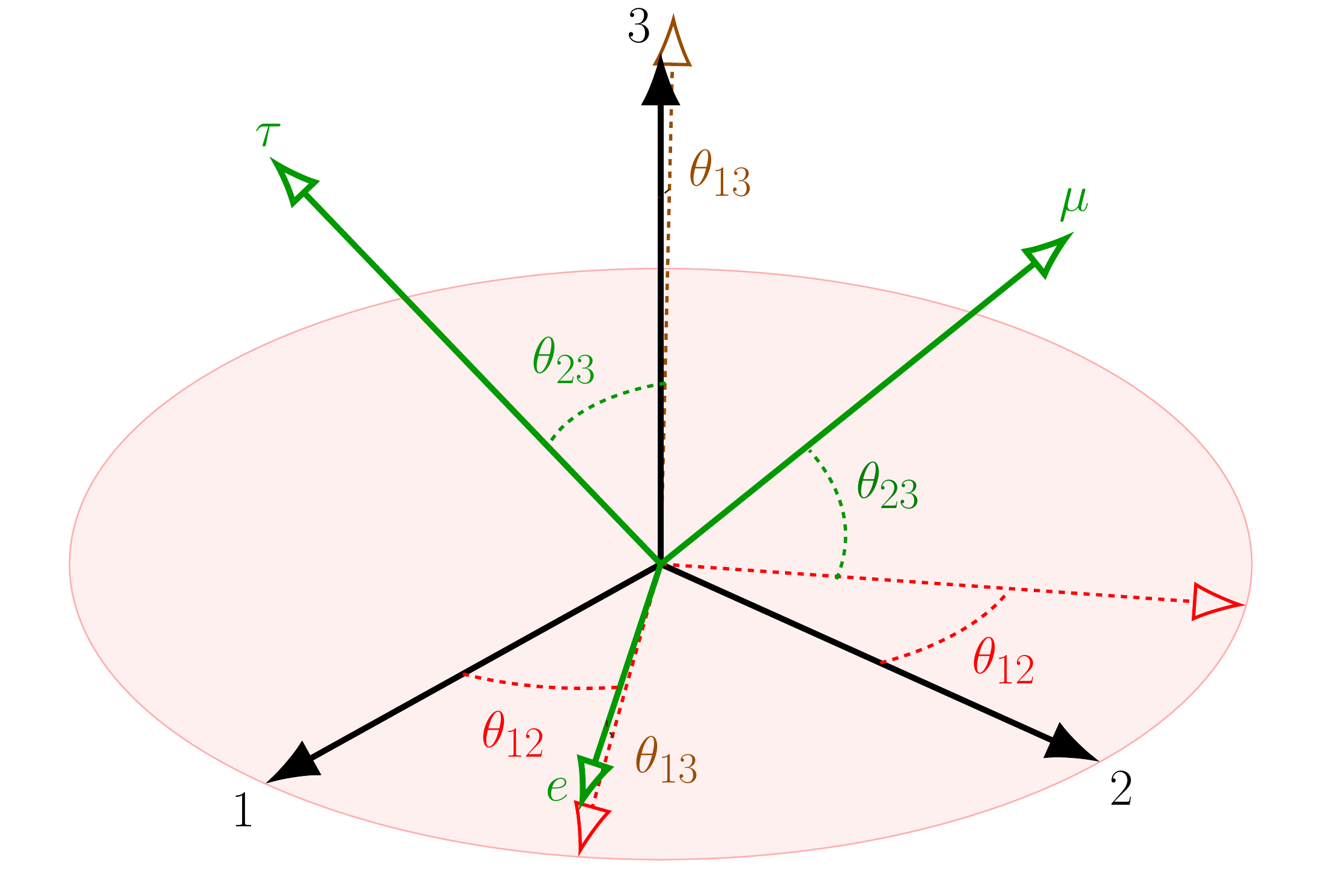}
\includegraphics[height=1.7in]{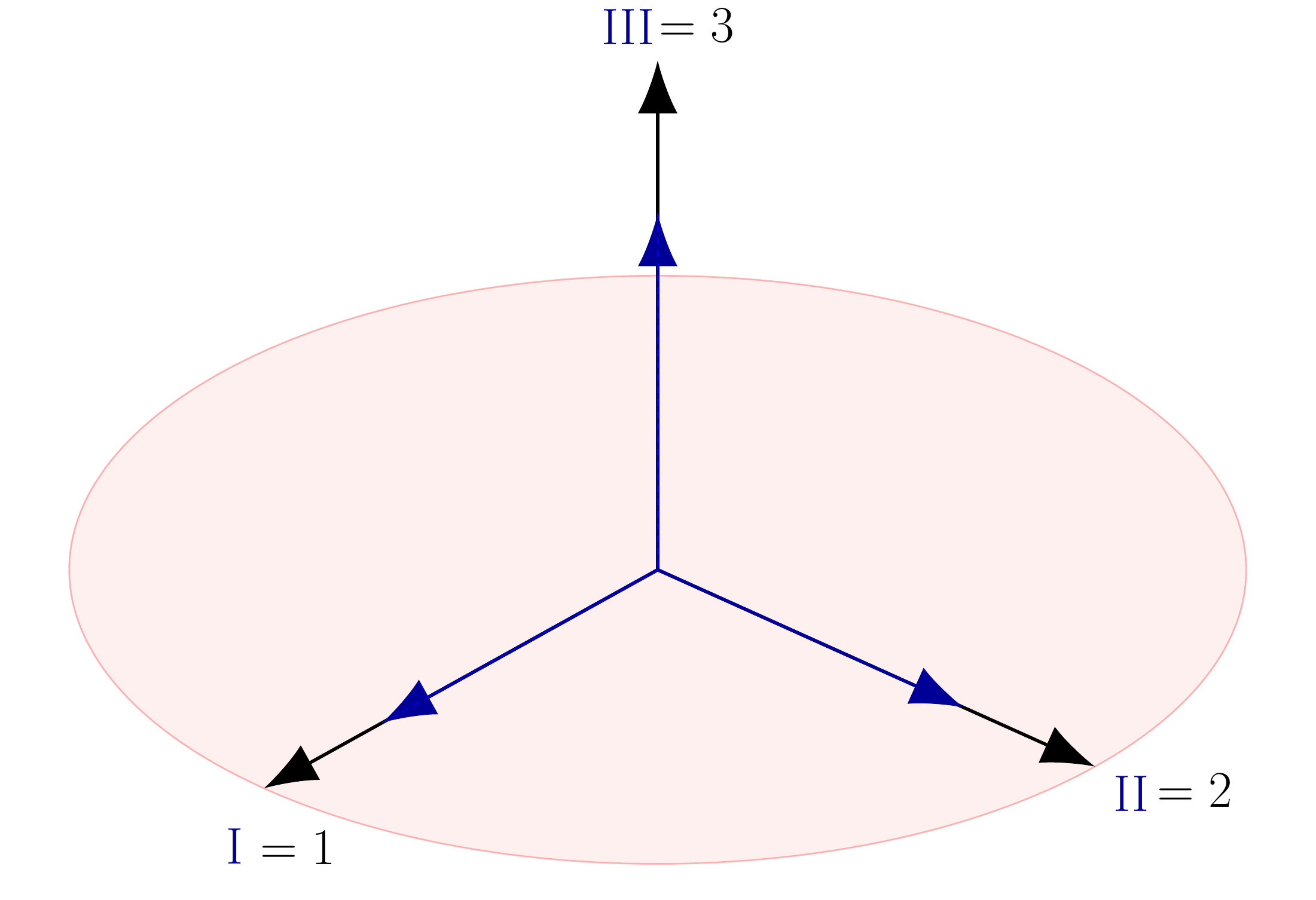}
\caption{Representation of different lepton bases in lepton flavour space.The light neutrino mass
eigenstates is used as reference basis. In the left panel the weak basis is shown with an indication 
of the three lepton mixing angles and in the right panel the heavy neutrino flavour basis is shown
in the simple case of form-dominance models, when it coincides with the light neutrino basis (from \cite{pdbrefsam}).}
\label{fig:magnet}
\end{figure}
If, as mentioned, one parameterises the orthogonal
matrix as the product of a rotation times a Lorentz boost in flavour space, this class of models
corresponds just to the simple trivial identity case (or a trivial permutation of light and heavy flavour states).
Interestingly, these models typically arise imposing a discrete flavour symmetry \cite{kingreview}. 
The bridging matrix $B$
of course also coincides with the permutation matrix in these models. As we will see, leptogenesis necessarily implies that some deviation from form-dominance is necessary to reproduce the observed baryon asymmetry of the universe. 

\subsubsection*{All mixing from the RH sector and $SO(10)$-inspired models}

The opposite limit case occurs when all leptonic mixing is generated by the RH sector,
implying  $V_L =I$.  In this case one still obtains analytical expressions for the 
RH neutrino masses, though not so simple as before, explicitly \cite{decrypting}
\be
M_1 =  {m^2_{D1} \over |m_{\nu ee}|}\, , \;\; 
M_2 = {m^2_{D2} \over m_1 \, m_2 \, m_3} \, {|m_{\nu ee}| \over |(m_{\nu}^{-1})_{\t\t}|} \, , \;\; 
M_3 =  m^2_{D3} \, |(m_{\nu}^{-1})_{\t\t}| \,   .
\ee
In the case of $SO(10)$-inspired models \cite{SO10inspired} 
one has $V_L \simeq V_{CKM} \simeq I$ and in addition
the Dirac neutrino masses are not too different from the up quark masses.
In this case the RH neutrino mass spectrum is highly hierarchical and 
when current neutrino mixing data are plugged into the expressions, barring
strongly fine tuned case of compact spectrum when both 
$|m_{\nu ee}|$ and $|(m_{\nu}^{-1})_{\t\t}|$ are very small, one has
$M_1 \ll 10^9 \, {\rm GeV}$, with important consequences for leptogenesis. 

An analytic expression for $U_R$ can be found in \cite{decrypting}, here it is interesting just to notice 
that in the hierarchical case, for $m_1 \rightarrow 0$, one has $U_R \rightarrow I$ and the expressions for the
heavy neutrino masses correctly tend to the form-dominance case. This might
sound like a paradox since we are assuming $V_L = I$ but the point is that in this case
the mixing is the result of a balance between numerator and denominator in the seesaw formula
and this is a perfectly viable case considering that this balance can well explain the observed large mixing
angles. Of course one can wonder whether this corresponds to a too special (though not fined-tuned) 
choice of parameters to be true but  it is still interesting that it naturally 
emerges as a perfectly viable solution within $SO(10)$-inspired models barring fine-tuning in the seesaw formula
and taking into account the experimental data.  Since in the exact limit all $C\!P$ asymmetries vanish, this might be seen as a reason why  one could expect a deviation from the hierarchical limit within these models.
We will see that indeed leptogenesis sets a strict lower bound on $m_1$ within $SO(10)$-inspired models.

\subsubsection*{Two right-handed neutrino models}

In the limit $M_3 \gg 10^{15}\,{\rm GeV}$, one necessarily has $m_1 \ll 10^{-5}\,{\rm eV}$
and in the seesaw formula the heaviest RH neutrino decouples and one obtains an effective
two RH neutrino formula with a great parameter reduction (from eighteen to only eleven) \cite{2RHnu}. 
These models are interesting because they are realised in different grand-unified models
combined with a discrete flavour symmetry (see for example \cite{bjorkeroth}).

However, these models can be also realised in the limit when one Yukawa coupling, or equivalently
one neutrino Dirac mass $m_{Di}$, vanishes. This is quite interesting since
in this case the decoupled heavy neutrino can have any mass and can be a candidate 
of dark matter \cite{ad}, as we discuss in Section 4.



\section{Leptogenesis}

We do not observe primordial antimatter in the universe. The baryon-to-photon ratio, measured
very precisely by cosmological observations finding \cite{planck2018}
\be
\eta_{B0} = (6.12 \pm 0.04) \times 10^{-10} \,  ,
\ee
can then be considered as a measurement of the baryon asymmetry of the universe that survived
the sequence of different particle species annihilations occurred in the early universe. 
A model of baryogenesis, where the asymmetry is generated dynamically after the inflationary stage, is 
considered as the most reasonable explanation. A successful model of baryogenesis 
cannot be found within the SM and, therefore, the baryon asymmetry of the universe  is 
regarded as a strong evidence of new physics.  Leptogenesis is a scenario of baryogenesis
relying on neutrino properties and after the discovery of neutrino masses and mixing it became
greatly attractive. Currently many versions exist but the minimal original one \cite{fy} still
remains not only viable but also the most appealing since it can be naturally embedded
within models of new physics such as grand-unified models and/or models of flavour or combinations. 

The minimal scenario of leptogenesis (see \cite{review} for a review) relies on the type-I
seesaw extension of the SM that we discussed in the previous section and on the assumption 
of thermal production of the heavy neutrinos, implying that the reheat temperature has
to be sufficiently high compared to the the mass of the RH neutrinos whose decay produce 
the contribution to the final asymmetry reproducing the measured one.  Heavy RH neutrino
decays produce a $B-L$ asymmetry that is injected in the form of lepton number. This is 
rapidly partly reprocessed by sphaleron non-perturbative processes into a baryon asymmetry
if the temperature of production is higher than the sphaleron freeze-out temperature  
($T_{\rm sph}^{\rm off} \simeq 132\,{\rm GeV}$ \cite{rummukainen}). The baryon-to-photon 
ratio predicted by leptogenesis can then be expressed in terms of the final $B-L$ asymmetry 
$N_{B-L}^{\rm fin}$ as $\eta_{B0}=a_{\rm sph}\,N_{B-L}^{\rm fin}/N_\g^{\rm rec}$, where
$a_{\rm sph} \simeq 1/3$ is the fraction of $B-L$ asymmetry  in the form of a baryon asymmetry. 

When flavour effects are taken into account \cite{flavoureffects,geometry,vives}
the final asymmetry, in general, given
by a sum of contributions both on heavy neutrino flavours and on charged lepton flavours. This crucially
depends on the RH neutrino mass spectrum and for this reason the requirement of 
successful leptogenesis imposes constraints on the RH neutrino mass spectrum. 
The most important one is that the mass of  RH neutrinos that dominantly produces the asymmetry, not necessarily the lightest ones, has to be higher than $\sim 10^9 \,{\rm GeV}$ if one considers a hierarchical RH neutrino spectrum and barring  fine-tuning in the seesaw formula. This lower bound also translates into a similar lower bound on $T_{\rm RH}$ \cite{bounds,geometry}.

If we go back to the case of models where all mixing stems from the LH sector, with $U = V^\dagger_L$,
and if one imposes a discrete flavour symmetry, then this typcially leads to 
$m_{D\star} \equiv m_{D1} = m_{D2} = m_{D3}$ and
one obtains a very simple heavy neutrino mass spectrum $M_{I}=m^2_{D\star}/m_i$. If in addition one also imposes  $m_1 \ll 10^{-5}\,{\rm eV}$, effectively the heaviest RH neutrino decouples and one can consider an effective two RH neutrino model. 
In this case the lower bound on the lightest RH neutrino is $\sim 10^{10}\,{\rm GeV}$ \cite{2RHnulep}.
The flavour symmetry needs to be broken to have non-vanishing $C\!P$ asymmetries 
\cite{jenkins,bdfn}. In these models one does not get, in general, definite predictions on the mixing 
parameters. These can be obtained within a specific model typically combining a discrete flavour symmetry
with grand-unification \cite{bjorkeroth2}.  

If we combine $SO(10)$-inspired models  with leptogenesis
($SO(10)$-inspired leptogenesis), then the predictive power greatly increases and one gets 
predictions also on the mixing parameters. The reason is that in that case, as anticipated,
since the asymmetry produced by the lightest RH neutrinos is negligible, this has necessarily
to be produced from next-to-lightest RH neutrinos, realising a so-called $N_2$-leptogenesis
scenario \cite{geometry,vives}. In this case one has also to worry that the wash-out from lightest
RH neutrinos is weak enough for an asymmetry in some flavour to survive prior to the 
freeze-out of sphalerons. This condition, combined with the requirement of successful leptogenesis, 
produces constraints on all low energy neutrino parameters \cite{riotto1,riotto2}. 
In particular a very robust constraint is given by a lower bound on the lightest neutrino mass
$m_1 \gtrsim 1\,{\rm meV}$ and it should also be said that, given the current upper bound 
on neutrino masses from cosmology, $SO(10)$-inspired leptogenesis
works only for normal ordering, as favoured by the latest data. 

An interesting feature of $SO(10)$-inspired leptogenesis is that, for a much more
restricted region of parameters, it can also be {\em strong}, meaning that the final asymmetry can be independent
of the initial conditions, and in particular large pre-existing asymmetries, in all three charged lepton flavours, can be efficiently washed-out. This requires quite a specific set of constraints on low energy neutrino parameters
whose full realisation could be basically interpreted as a signature. For example, the lightest neutrino mass
has to be comprised within quite a narrow range of values, $m_1 \simeq (10$--$30)\,{\rm meV}$ (the exact range depends on how large is the pre-existing asymmetry to be washed-out, see \cite{STSO10} for details)  
that starts to be right now to be tested by cosmological observations (see footnote at page 2). 
Also, quite interestingly, the solution successfully predicted a non-vanishing value of $\theta_{13}$.
Another important feature of {\em strong thermal $SO(10)$-inspired leptogenesis} is that
it can be hardly compatible with the atmospheric neutrino mixing angle in the second octant \cite{STSO10atm},
as currently slightly favoured by global analyses. In the next years it will certainly very interesting  whether
new experimental results will further support this solution or  rule it out.

It should be noticed that $SO(10)$-inspired conditions can be also realised also not necessarily $SO(10)$ models. 
For example in \cite{flavourcoupling} it has been shown that a model based on a combination of a
$A_4$ discrete flavour symmetry with Pati-Salam grand-unified group, leads to $SO(10)$-inspired conditions
and one can reproduce lepton parameters and also obtain successful $N_2$ leptogenesis. Interestingly, in this case the atmospheric neutrino mixing angle needs to be in the second octant as favoured by latest global analyses.
If one wants also to get in addition a realistic fit of quark parameters, this has been found within
a $SO(10) \times S_4 \times \mathbb{Z}_4^R \times \mathbb{Z}_4^3$ model,
where $\mathbb{Z}_4^R$ is an R symmetry while the other three $\mathbb{Z}_4$'s symmetries are 
shaping symmetries \cite{perdomo}.

\section{Dark matter}

Cosmological observations measure with great precision the abundance of cold dark matter in the universe.
The latest results from the {\em Planck} collaboration find \cite{planck2018} 
$\O_{\rm C DM}\,h^2 = 0.11933 \pm 0.00091$. 
Can the simple type-I seesaw Lagrangian (\ref{seesawmasster}) also address the dark matter puzzle?
In this case one of the three heavy RH neutrinos should play the role of dark matter particle. 
A solution is obtained, the so-called $\nu$MSM model \cite{nuMSM}, if the lightest RH neutrino mass is much smaller than the electron mass in a way that
the dominant decay channel is into three neutrinos and the rate can be so strongly suppressed to have
a life-time much longer with the age of the universe.  This points to a mass of the dark matter RH neutrino of order of keV that is
interesting since it would behave as warm dark matter, potentially 
able to solve some claimed problems in the large scale structure at scales of the galactic sub-halos. 
The dark matter RH neutrino would be produced by the mixing with the LH neutrinos and interestingly the 
correct abundance can be produced while at the same time the seesaw formula can satisfy the experimental
results from neutrino mixing experiments. However dark matter RH neutrinos would also sub-dominantly decay
radiatively and X-ray constraints right now exclude a non-resonance production from the  mixing. One has then
to introduce further ingredients in the picture that becomes much more contrived.
Moreover, in the $\nu$MSM, neutrino Yukawa couplings are still  much smaller than those of other massive fermions so that one of the original motivation of the seesaw mechanism is actually not addressed. 

An alternative solution, with values of the RH neutrino masses above the TeV scale implying higher
neutrino Yukawa couplings, is to consider one of the three RH neutrinos decoupled and stable. 
This implies that its Yukawa couplings have basically to vanish and this can be justified imposing  some symmetry.
In order to produce the dark matter RH neutrino however it is necessary to introduce some new interaction. 
An attractive option is to consider the existence of new interactions described by the the 5-dim non renormalizable  operator \cite{ad}
\be\label{anisimov}
{\cal O}_A = {\lambda_{IJ} \over \L} \, \Phi^\dagger \, \Phi \, \overline{N_I^c} \, N_J \,  ,
\ee
inducing a RH-RH neutrino mixing able to produce via non-adiabatic resonant conversions a 
RH neutrino DM abundance. The same operator is eventually also responsible also for the decays
of the dark matter RH neutrinos and this implies both a lower bound and an upper bound on the mass
singling out a window within $100\,{\rm TeV}$--$10\,{\rm PeV}$ that is quite interesting since it implies 
some contribution to the high energy neutrino flux that is now detected by the IceCube neutrino telescope,  providing a way to test the mechanism.  Interestingly, the other two coupled RH neutrino decays can also
reproduce the observed baryon asymmetry via leptogenesis \cite{unified}.  In this way one realises
a unified picture of neutrino masses, leptogenesis and dark matter testable at neutrino telescopes.

In conclusion a solution to the cosmological puzzles of matter-antimatter asymmetry and dark matter
of the universe related to neutrino properties is not only possible but also an attractive possibility that will be tested  during next years  at neutrino telescopes.

\Acknowledgements
It is a great pleasure to thank the organisers for such an interesting conference. 
I also wish to thank Marco Chianese, Kareem Farrag, Michele Re Fiorentin, Teppei Katori, 
Sergio Palomares-Ruiz, Rome Samanta, Ye-Ling Zhou for a fruitful collaboration on the topics discussed.


\begin{thebibliography}{99}

\bibitem{nufit2018}
  I.~Esteban, M.~C.~Gonzalez-Garcia, A.~Hernandez-Cabezudo, M.~Maltoni and T.~Schwetz,
 {\em Global analysis of three-flavour neutrino oscillations: synergies and tensions in the 
 determination of $\theta_{23}, \delta_{CP}$, and the mass ordering},
  JHEP {\bf 1901} (2019) 106
  [arXiv:1811.05487 [hep-ph]].

\bibitem{planck2015}
P.~A.~R.~Ade {\it et al.} [Planck Collaboration],
  {\em Planck 2015 results. XIII. Cosmological parameters},
  Astron.\ Astrophys.\  {\bf 594} (2016) A13
  [arXiv:1502.01589].


\bibitem{planck2018}
Y.~Akrami {\it et al.} [Planck Collaboration],
  {\em Planck 2018 results. I. Overview and the cosmological legacy of Planck},
  arXiv:1807.06205 [astro-ph.CO];
See also talk by M. Lattanzi at this conference.

\bibitem{arkani}
N.~Arkani-Hamed, S.~Dimopoulos, G.~R.~Dvali and J.~March-Russell,
  {\em Neutrino masses from large extra dimensions},
  Phys.\ Rev.\ D {\bf 65} (2001) 024032
  [hep-ph/9811448].

\bibitem{neubert}
 Y.~Grossman and M.~Neubert,
  {\em Neutrino masses and mixings in nonfactorizable geometry},
  Phys.\ Lett.\ B {\bf 474} (2000) 361
  [hep-ph/9912408].

\bibitem{lindner}
K.~Dick, M.~Lindner, M.~Ratz and D.~Wright,
  {\em Leptogenesis with Dirac neutrinos},
  Phys.\ Rev.\ Lett.\  {\bf 84} (2000) 4039
  [hep-ph/9907562].

\bibitem{seesaw}
P.~Minkowski,
  {\em $\mu \to e \gamma$ At A Rate Of One Out Of 1-Billion Muon Decays?},
  Phys.\ Lett.\  B {\bf 67} (1977) 421;
T. Yanagida, {\em Horizontal gauge symmetry and masses of neutrinos},
in Proceedings of the Workshop on Unified Theory and Baryon Number
of the Universe, eds. O. Sawada and A. Sugamoto (KEK, 1979) p.95;
  P.~Ramond, 
Invited talk given at Conference: C79-02-25
(Feb 1979) p.265-280, CALT-68-709,
  {\em The Family Group in Grand Unified Theories},
  hep-ph/9809459;
 M.~Gell-Mann, P.~Ramond and R.~Slansky,
  {\em Complex Spinors and Unified Theories},
  Conf.\ Proc.\ C {\bf 790927} (1979) 315
  [arXiv:1306.4669 [hep-th]];
R.~Barbieri, D.~V.~Nanopoulos, G.~Morchio and F.~Strocchi,
  {\em Neutrino Masses in Grand Unified Theories},
  Phys.\ Lett.\  {\bf 90B} (1980) 91;
  R.~N.~Mohapatra and G.~Senjanovic,
  {\em Neutrino Mass and Spontaneous Parity Nonconservation},
  Phys.\ Rev.\ Lett.\  {\bf 44} (1980) 912.


\bibitem{casasibarra}
J.~A.~Casas and A.~Ibarra,
  {\em Oscillating neutrinos and $\mu \rightarrow e + \gamma$},
  Nucl.\ Phys.\ B {\bf 618} (2001) 171
  [hep-ph/0103065].

\bibitem{pdbrefsam}
  P.~Di Bari, M.~Re Fiorentin and R.~Samanta,
  {\em Representing seesaw neutrino models and their motion in lepton flavour space},
  arXiv:1812.07720 [hep-ph].

\bibitem{geometry}
 P.~Di Bari, {\em Seesaw geometry and leptogenesis},
  Nucl.\ Phys.\ B {\bf 727} (2005) 318
  [hep-ph/0502082].

\bibitem{chenking}
M.~C.~Chen and S.~F.~King,
  JHEP {\bf 0906} (2009) 072
  [arXiv:0903.0125 [hep-ph]].

\bibitem{kingreview}
S.~F.~King,  {\em Unified Models of Neutrinos, Flavour and CP Violation},
  Prog.\ Part.\ Nucl.\ Phys.\  {\bf 94} (2017) 217
  [arXiv:1701.04413 [hep-ph]].
  
\bibitem{decrypting}
P.~Di Bari, L.~Marzola and M.~Re Fiorentin,
 {\em Decrypting $SO(10)$-inspired leptogenesis},
  Nucl.\ Phys.\ B {\bf 893} (2015) 122
  [arXiv:1411.5478 [hep-ph]].

\bibitem{SO10inspired}
A.~Y.~Smirnov,
  {\em Seesaw enhancement of lepton mixing},
  Phys.\ Rev.\ D {\bf 48} (1993) 3264
  [hep-ph/9304205];
W.~Buchmuller and M.~Plumacher,
  {\em Baryon asymmetry and neutrino mixing},
  Phys.\ Lett.\ B {\bf 389} (1996) 73 [hep-ph/9608308];
E.~Nezri and J.~Orloff,
  {\em Neutrino oscillations versus leptogenesis in $SO(10)$ models},
  JHEP {\bf 0304} (2003) 020
  [hep-ph/0004227];
F.~Buccella, D.~Falcone and F.~Tramontano,
  {\em Baryogenesis via leptogenesis in $SO(10)$ models},
  Phys.\ Lett.\ B {\bf 524} (2002) 241 [hep-ph/0108172];
G.~C.~Branco, R.~Gonzalez Felipe, F.~R.~Joaquim and M.~N.~Rebelo,
  {\em Leptogenesis, CP violation and neutrino data: What can we learn?},
  Nucl.\ Phys.\ B {\bf 640} (2002) 202 [hep-ph/0202030];
E.~K.~Akhmedov, M.~Frigerio and A.~Y.~Smirnov, JHEP {\bf 0309}, 021 (2003).


\bibitem{2RHnu}
 S.~F.~King,
{\em Large mixing angle MSW and atmospheric neutrinos from single  RH
 neutrino dominance and $U(1)$ family symmetry},
  Nucl.\ Phys.\  B {\bf 576} (2000) 85  [arXiv:hep-ph/9912492];
P.~H.~Frampton, S.~L.~Glashow and T.~Yanagida,
 {\em Cosmological sign of neutrino CP violation},
  Phys.\ Lett.\  B {\bf 548} (2002) 119
  [arXiv:hep-ph/0208157].
P.~H.~Chankowski and K.~Turzynski,
  {\em Limits on $T_{\rm reh}$ for thermal leptogenesis with hierarchical neutrino masses},
  Phys.\ Lett.\  B {\bf 570} (2003) 198
  [arXiv:hep-ph/0306059];
A.~Ibarra and G.~G.~Ross,
  {\em Neutrino phenomenology: The case of two right handed neutrinos},
  Phys.\ Lett.\  B {\bf 591} (2004) 285.

\bibitem{bjorkeroth}
 F.~Bj\"{o}rkeroth, F.~J.~de Anda, I.~de Medeiros Varzielas and S.~F.~King,
  {\em Leptogenesis in minimal predictive seesaw models},
  JHEP {\bf 1510} (2015) 104
  [arXiv:1505.05504 [hep-ph]].

\bibitem{ad}
A.~Anisimov and P.~Di Bari,
 {\em Cold Dark Matter from heavy Right-Handed neutrino mixing},
  Phys.\ Rev.\ D {\bf 80} (2009) 073017
  [arXiv:0812.5085 [hep-ph]].

\bibitem{fy}
M.~Fukugita and T.~Yanagida,
  {\em Baryogenesis Without Grand Unification},
  Phys.\ Lett.\ B {\bf 174} (1986) 45.

\bibitem{review}
S.~Blanchet and P.~Di Bari,
  {\em The minimal scenario of leptogenesis},
  New J.\ Phys.\  {\bf 14} (2012) 125012
  [arXiv:1211.0512 [hep-ph]].




\bibitem{rummukainen}
M.~D'Onofrio, K.~Rummukainen and A.~Tranberg,
  {\em Sphaleron Rate in the Minimal Standard Model},
  Phys.\ Rev.\ Lett.\  {\bf 113} (2014) no.14,  141602
  [arXiv:1404.3565].

\bibitem{flavoureffects}
R.~Barbieri, P.~Creminelli, A.~Strumia and N.~Tetradis,
  {\em Baryogenesis through leptogenesis},
  Nucl.\ Phys.\ B {\bf 575} (2000) 61
  [hep-ph/9911315];
 A.~Abada, S.~Davidson, F.~X.~Josse-Michaux, M.~Losada and A.~Riotto,
  JCAP {\bf 0604} (2006) 004
  doi:10.1088/1475-7516/2006/04/004
  [hep-ph/0601083];
E.~Nardi, Y.~Nir, E.~Roulet and J.~Racker,
  JHEP {\bf 0601} (2006) 164
  doi:10.1088/1126-6708/2006/01/164
  [hep-ph/0601084].


\bibitem{bounds}
S.~Davidson and A.~Ibarra,
  {\em A Lower bound on the right-handed neutrino mass from leptogenesis},
  Phys.\ Lett.\ B {\bf 535} (2002) 25
  [hep-ph/0202239];
W.~Buchmuller, P.~Di Bari and M.~Plumacher,
  {\em Cosmic microwave background, matter - antimatter asymmetry and neutrino masses},
  Nucl.\ Phys.\ B {\bf 643} (2002) 367
   Erratum: [Nucl.\ Phys.\ B {\bf 793} (2008) 362]
  [hep-ph/0205349];
S.~Blanchet and P.~Di Bari,
  {\em Flavor effects on leptogenesis predictions},
  JCAP {\bf 0703} (2007) 018
  [hep-ph/0607330];
S.~Blanchet and P.~Di Bari,
  {\em New aspects of leptogenesis bounds},
  Nucl.\ Phys.\ B {\bf 807} (2009) 155
  [arXiv:0807.0743 [hep-ph]].


\bibitem{2RHnulep}
S.~Antusch, P.~Di Bari, D.~A.~Jones and S.~F.~King,
  {\em Leptogenesis in the Two Right-Handed Neutrino Model Revisited},
  Phys.\ Rev.\ D {\bf 86} (2012) 023516
  [arXiv:1107.6002 [hep-ph]].

\bibitem{jenkins}
E.~E.~Jenkins and A.~V.~Manohar,
  {\em Tribimaximal Mixing, Leptogenesis and $\theta_{13}$},
  Phys.\ Lett.\ B {\bf 668} (2008) 210
  [arXiv:0807.4176 [hep-ph]].
  
\bibitem{bdfn}
  E.~Bertuzzo, P.~Di Bari, F.~Feruglio and E.~Nardi,
  {\em Flavor symmetries, leptogenesis and the absolute neutrino mass scale},
  JHEP {\bf 0911} (2009) 036
  [arXiv:0908.0161].


\bibitem{bjorkeroth2}
See for example F.~Bj\"{o}rkeroth, F.~J.~de Anda, I.~de Medeiros Varzielas and S.~F.~King,
  {\em Leptogenesis in a $\Delta(27) \times SO(10)$ SUSY GUT},
  JHEP {\bf 1701} (2017) 077
  [arXiv:1609.05837 [hep-ph]].



\bibitem{vives}
O.~Vives,
{\em Flavor dependence of CP asymmetries and thermal leptogenesis with strong right-handed neutrino mass hierarchy},
  Phys.\ Rev.\ D {\bf 73} (2006) 073006
  [hep-ph/0512160].




\bibitem{riotto1}
P.~Di Bari and A.~Riotto,
  {\em Successful type I Leptogenesis with SO(10)-inspired mass relations},
  Phys.\ Lett.\ B {\bf 671} (2009) 462
  [arXiv:0809.2285 [hep-ph]].

\bibitem{riotto2}
 P.~Di Bari and A.~Riotto,
  {\em Testing SO(10)-inspired leptogenesis with low energy neutrino experiments},
  JCAP {\bf 1104} (2011) 037
  [arXiv:1012.2343 [hep-ph]].

\bibitem{STSO10}
P.~Di Bari and L.~Marzola,
 {\em SO(10)-inspired solution to the problem of the initial conditions in leptogenesis},
  Nucl.\ Phys.\ B {\bf 877} (2013) 719
  [arXiv:1308.1107].

\bibitem{STSO10atm}
M.~Chianese and P.~Di Bari,
{\em Strong thermal $SO(10)$-inspired leptogenesis in the light of recent results from long-baseline neutrino experiments}, JHEP {\bf 1805} (2018) 073
  [arXiv:1802.07690 [hep-ph]].

\bibitem{flavourcoupling}
P.~Di Bari and S.~F.~King,
 {\em Successful $N_2$ leptogenesis with flavour coupling effects in realistic unified models},
  JCAP {\bf 1510} (2015) no.10,  008
  [arXiv:1507.06431].


\bibitem{perdomo}
F.~J.~de Anda, S.~F.~King and E.~Perdomo,
 {\em $\mathbf{SO(10)}\times \mathbf{S_4}$ grand unified theory of flavour and leptogenesis},
  JHEP {\bf 1712} (2017) 075
   Erratum: [JHEP {\bf 1904} (2019) 069]
  [arXiv:1710.03229 [hep-ph]].


\bibitem{nuMSM}
T.~Asaka, S.~Blanchet and M.~Shaposhnikov,
 {\em The $\nu$MSM, dark matter and neutrino masses},
  Phys.\ Lett.\ B {\bf 631} (2005) 151
  [hep-ph/0503065].

\bibitem{unified}
P.~Di Bari, P.~O.~Ludl and S.~Palomares-Ruiz,
 {\em Unifying leptogenesis, dark matter and high-energy neutrinos with right-handed neutrino mixing via Higgs portal},
  JCAP {\bf 1611} (2016) no.11,  044
  [arXiv:1606.06238 [hep-ph]].

\end{thebibliography}
\end{document}